\newcommand\z{\zeta}

\renewcommand\k{\kappa}

\newcommand\p{\pi}
\newcommand\s{\sigma}

\renewcommand\o{\omega}

\newcommand\sech{\mathrm{sech}}

\newcommand\ra{\rightarrow}

\newcommand{\SCR}{Schr\"odinger~}
\newcommand{\GP}{Gross-Pitaevskii~}
\newcommand{\diracslash}[1]{#1\llap{/\kern2pt}}

\newcommand{\be}{\begin{equation}}
\newcommand{\ee}{\end{equation}}
\newcommand{\bea}{\begin{eqnarray}}
\newcommand{\eea}{\end{eqnarray}}
\newcommand{\ba}[1]{\begin{array}{#1}}
\newcommand{\ea}{\end{array}}

\documentclass[12pt]{iopart}
\usepackage{subfigure}
\usepackage{graphicx}
\begin{document}
\title{Controlling Grey Solitons in a Trap}

\author{Ayan Khan}
 \address{ Department of Physics, University of Camerino, 62032, Italy}
\ead{ayan.khan@unicam.it}
 \author{Rajneesh Atre}
 \address{Department of Physics, Jaypee  Institute of Engineering and Technology, 
Guna, 473226, India} 
\author{Prasanta K. Panigrahi}
\address{Indian Institute of Science Education and Research, Kolkata, 741252, India}
\ead{prasanta@prl.res.in}

\date{\today}

\begin{abstract}
Experimentally observed grey solitons are analytically extracted from a physically viable \GP equation.
Associated Lieb and Bogoliubov modes are calculated for these class of solitons. It is observed that, these nonlinear excitations have strong coupling with the trap at low momenta and hence can be effectively isolated from the Bogoliubov sound modes, which responds weakly to harmonic confinement. This strong coupling with the trap also makes the grey soliton amenable for control and manipulation through both trap modulation and temporal variation of the two-body interaction.
\end{abstract}

\submitto{\jpb}
\maketitle 
\section{Introduction}
Recent observation of oscillations between stable localized grey solitons in a cigar shaped 
Bose-Einstein condensate (BEC) and vortex rings \cite{levy}  
has led to significant interest regarding interaction between the collective
matter wave excitations. These coherent structures are relevant for atom interferometry, making their 
control and manipulation an area of active research \cite{levy,gross,becker,sol}. The nonlinearity of \GP (GP) equation,
describing BEC at the mean-field level, leads to these excitations, which have been explored extensively 
in recent times \cite{carr,ldcarr,jackson, komineas}.
The GP equation reduces to the well known nonlinear \SCR equation (NLSE) for a cigar-shaped BEC, which is an integrable model admitting stable soliton solutions \cite{zak}. Experimental observations of dark \cite{burger,deng} and bright \cite{khaykovich,strecker,khawaja,Wieman} solitons   
have led to considerable interest in understanding the nature of stable collective excitations of BEC in different spatial dimensions. A systematic study of the decay of dark soliton has also been carried out recently \cite{adams}.
The above mentioned grey solitons are produced through collision of two BECs, when the collisional energy is reduced to a level, where it is comparable with the interaction energy and the wavelength of the laser responsible for density modulation is larger than the healing length. In this scenario, nonlinearity plays a stronger role than dispersion. 
 
In 1963, in a second quantized formulation, Lieb discovered a collective 
excitation of the condensed bosons, which exhibited a periodic dispersion \cite{lieb}.
The same was later identified with a complex soliton at the mean field level \cite{kulish}. 
As compared to the dark soliton, which necessarily passes through the normal phase, with
zero condensate density, the Lieb mode can attain the asymptotic density, without reaching the local potential maximum corresponding to the normal phase. This is possible because of the complex envelope of the Lieb mode, analogous to the Bloch solitons in condensed matter systems. Interestingly this non-linear mode has a periodic dispersion very different from other collective excitations. Laboratory confirmation of this mode has been hampered by the fact that in the experimentally accessible low momenta regime, Lieb mode dispersion lies below that of the Bogoliubov excitations \cite{jackson,komineas,jack,pop,kom,ajack}. The instability of BEC for higher momentum values, where these two modes differ significantly excluded this domain from observation. In the recent years considerable attention is paid to the study of sound modes not only in Bose gases but also in Fermi systems \cite{recati,taylor}, whereas a systematic study with the new experimental developments on the solitonic mode is still lagging.

In this letter, we establish a completely analytical approach to achieve grey soliton solution in presence of a modulating harmonic trap and Feshbach induced scattering length. Further, a detailed study of their energy and momentum is reported. Explicit calculation reveals that, harmonic confinement  significantly affects the energy momentum profile at low momenta. The fact that, the trap does not allow a uniform density and the complex soliton's asymptotic behavior matches with the uniform state, leads to this strong interaction between grey solitons and the trap geometry. In contrast, the Bogoliubov mode couples weakly with the trap. This opens up the possibility  for the coherent control and manipulation of these solitons at low energy, through the temporal modulation of the scattering length, as well as the trap \cite{moores,longhi,kruglov,liang,atre,engels,radha,alk}. We demonstrate this explicitly, by obtaining the self similar grey soliton profile in the presence of the harmonic confinement, with time dependent scattering length. The solitons are necessarily chirped and can be accelerated, compressed or brought to rest. The effect of the trap on the energy and momentum of the solitons is analytically computed.
\section{Complex Envelop Soliton}
At low temperature, the BEC, confined in an asymmetric trap, 
$V(x,y)=V_{0}+V_{1}=m\omega_{\perp}^{2}(x^2+y^2)/2+m\omega_{0}^{2}z^2/2$, can be made effectively one dimensional, if $\o_{\perp}>>\o_{0}$. The mean field GP equation, with $U=4\pi\hbar^2a_{s}/m$,
\bea
i\hbar\partial_{t}\Psi=(-\frac{\hbar^2}{2m}\vec{\nabla}^2+U|\Psi|^2+V)\Psi,
\eea
then allows separation of transverse and longitudinal dynamics, leading to the factorization of the order parameter, such that $\Psi=f(z,t)g(x,y,\s)$, where $\s$ is the local density of particles per unit length \cite{jack,ol,sala}.
In the weak coupling limit, $|g|^2$ has a Gaussian form and $f(z,t)$ satisfies \cite{jackson,komineas},
\bea
i\hbar\partial_{t}f(z,t)=-(\hbar^2/2m)\partial_{z}^2f+\hbar\o_{\perp}(1+2a_{s}|f|^2)f(z,t).\nonumber
\eea 
The Lieb mode has been identified with the complex envelope soliton solution: $|f(z)|^2=\s_{0}-\s_{0}\cos^2\theta\sech^2(z\cos\theta/\zeta)$ \cite{jackson,komineas}, where $\zeta$ is the coherence length and $\theta$ is the Mach angle. These modes correspond to localized density dips, which asymptotically match with the density $\s_{0}$ of the uniform BEC. Explicitly, $\theta=\sin^{-1}\frac{u}{c_{s}}$, which makes it evident that the soliton velocity $u$ is bounded by the sound speed $c_{s}$, where $c_{s}=\sqrt{\frac{2\hbar\o_{\perp}\s_{0}a_{s}}{m}}$. It is also clear that, the shallower solitons move faster. At $\theta=0$, where $u=0$, the grey soliton attains its maximum depth; at $\theta=\p/2$, $u=c_{s}$, the maximum attainable speed, where $\s(z,t)=\s_{0}$, corresponding to the uniform BEC.
The energy of this extended object is analytically computable, with respect to the constant background and is given by, 
$\frac{4}{3}E_{0}(1-\frac{u^2}{c_{s}^2})^{3/2}$, where $E_{0}=\sqrt{2\s_{0}a}\hbar\o_{\perp}\s_{0}a_{\perp}$. 

As will be seen, the presence of the trap strongly influences the grey solitons. This arises from the fact that, harmonic confinement does not allow an uniform density and the complex soliton's asymptotic behavior matches with the uniform state. This effect is particularly prominent in the low momenta regime, where the Bogoliubov modes are not significantly affected. In the following, we exactly obtain the self similar grey soliton profile in the presence of the harmonic confinement, a time dependent scattering length and a phenomenological loss term. This transparently reveals the grey soliton's response to the temporal changes of the trap and other parameters. The appropriately scaled GP equation, in dimensionless units, can be written in the form \cite{liang,atre}, 

\be \label{NLSE}
i\partial_{t}\psi=\Big(-\frac{1}{2}\partial_{zz}+\gamma(t)|\psi|^2
+\frac{1}{2}M(t)z^2+i\frac{g(t)}{2}-\frac{\nu(t)}{2}\Big)\psi. \nonumber\\\ee
Here, the interaction strength and the spring constant are 
$\gamma(t)=2a_{s}(t)/a_B$, $M(t)=\omega_{0}^{2}(t)/\omega_{\perp}^{2}$, respectively. 
Further $a_{\perp}=(\hbar/m\omega_{\perp})^{1/2}$ and $a_B$ is the Bohr radius. 

The ansatz solution is taken in the form, 
$\psi(z,t)=U(z,t)\exp[i\Phi(z,t)]$,
where,
\bea 
U(z,t)=B(t)\sqrt{\sigma(T)} \exp[i\chi(T)+G(t)/2],\nonumber
\eea
with $T=A(t)(z-l(t))$ and $G(t)=\int_{0}^{t}{g(t')dt'}$. The phase has a quadratic form exhibiting chirping: 
$\Phi(z,t)=a(t)+b(t)z-\frac{1}{2}c(t)z^2$.

The consistency conditions lead to a Riccati type equation, $c_{t}-c^{2}(t)=M(t)$ \cite{atre}. The width $A(t)$ is given by
$A(t)=A_{0}\exp\big({\int_{0}^{t}c(t')dt'}\big)$; further $\gamma(t)=\gamma_{0}e^{-G}A(t)/A_{0}$, $b(t)=A(t)$ and 
$\nu(t)=2 A^{2}\mu$. The center of mass motion is governed by,
\bea
l_{t}+cl-b=Au=V.\nonumber
\eea
It can be controlled through the trap, condensate motion and scattering length.
The amplitude is related to the width as, $B(t)=\sqrt{A(t)}$.
Explicit calculation shows, $a(t)=a_{0}-\frac{1-\bar{\mu}}{2}\int_{0}^{t}A^{2}(t')dt'$.
The current conservation yields, $\frac{\partial\chi}{\partial T}=u(1-\frac{\sigma_{0}}{\sigma})$, with boundary condition, $\chi'\ra 0$ for $\sigma\ra\sigma_{0}$,
where $\sigma_{0}$ is the equilibrium density of atoms in the moving frame $T$. Here $\bar{\mu}=-2\kappa\sigma_{0}=\mu+\lambda$. $\mu$, $\lambda$ are the chemical potential and constant parameter controlling the 
energy of excitation \cite{priyam} and $\kappa=\frac{\gamma_{0}}{A_{0}}$.
In this frame, the density equation can be cast in the convenient form \cite{jackson},
\begin{eqnarray}\label{hydro2}
\Bigg(\frac{\partial \sqrt{\sigma}}{\partial T}\Bigg)^{2}&=&(\kappa \sigma -u^{2})
\frac{(\sigma-\sigma_{0})^{2}}{2\sigma} \textrm{,}
\end{eqnarray}
\begin{figure}
\begin{center}
\includegraphics[width=3.5 in]{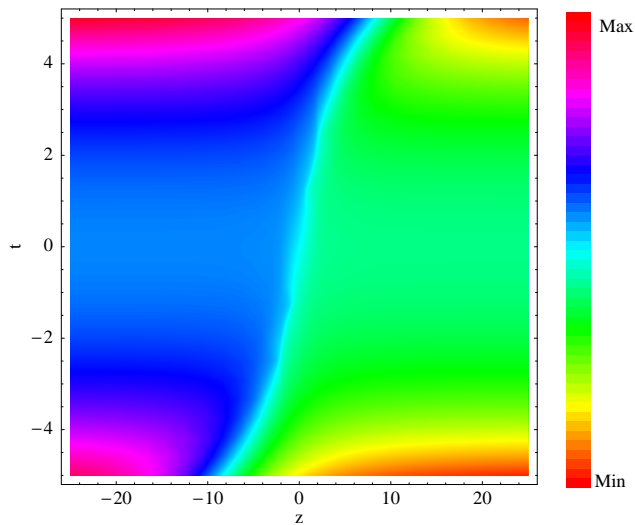}
\caption{Phase step of $\pi$ in grey soliton. The parameter values are $\s_{0}=10^7, \kappa=0.1, 
A_{0}=0.5, \theta=\pi/16$.}\label{cphase}
\end{center}
\end{figure}

The solution for density takes the self similar form \cite{khan}, 
\begin{eqnarray}\label{solution}
\sigma(z,t)=\sigma_{0}-\sigma_{0}\cos^{2}{\theta}\,\, 
\sech^{2}\,{[\frac{A(z-l)\cos{\theta}}{\zeta}]} \textrm{,}
\end{eqnarray}
where the Mach angle in the new frame is given by,
\bea\label{mach}
\theta=\sin^{-1}\frac{V}{C_{s}}=\sin^{-1}\frac{u}{c_{s}}.
\eea
In the above analysis the temporal variation of the background modifies the sound velocity: 
$C_{s}=A\sqrt{\kappa\sigma_{0}}=Ac_{s}$ \cite{sergei}.

The phase step is a direct attribute of the complex soliton, the obtained theoretical value is $\pi$ \cite{carr}, whereas the experimental observation yields $0.7\pi$ \cite{levy}. The phase change across the soliton profile as a function of space-time is depicted in Fig. (\ref{cphase}). 
\begin{figure}
\begin{center}
\includegraphics[width=3.5 in]{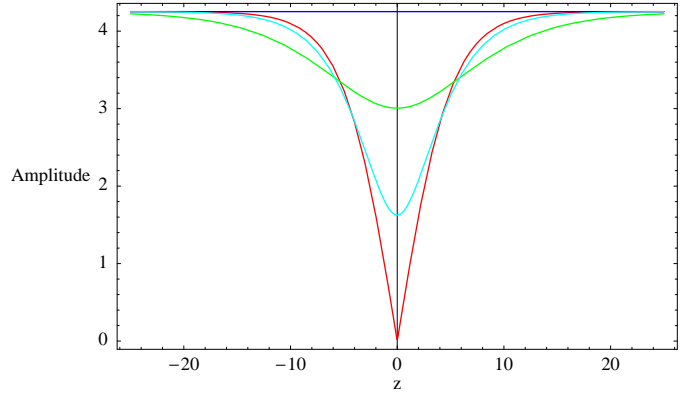}
\caption{Decrease of soliton depth with increasing velocity in a harmonic trap, 
calculated for $\theta=0,\pi/8,\pi/4$ and $\pi/2$, where 
$\theta=\sin^{-1}\frac{V}{C_{s}}$, $V$ being the soliton velocity and $C_{s}$ the speed of sound.}\label{solvel}
\end{center}
\end{figure}

As in the case of the grey soliton without a trap, in the present case also, the shallower solitons move faster than the deeper ones, as depicted in Fig. (\ref{solvel}). Higher $\theta$ values correspond to higher speed as is evident from Eq.(\ref{mach}) and correspond to shallower grey solitons.
\section{Energy Momentum Dispersion}
The energy of soliton per unit length
can be evaluated by subtracting the background energy such that $E=W-W_{0}$, with, 
\bea 
W&=&\int[\frac{1}{2}(\frac{\partial \psi^{*}}{\partial z}
\frac{\partial \psi}{\partial z})+\frac{1}{2}\gamma(t)(\psi^{*}\psi)^{2})+
\frac{1}{2}M(t)z^{2}\psi^{*}\psi]dz \nonumber\eea
and $\psi_{0}=\rho_{0}e^{i\Phi(z,t)+G/2}$, which corresponds to the BEC with a trivial phase in a trap of similar parameter values. The computed energy,
\bea
E&=&e^{G}(\frac{4}{3}\kappa A^{2}\sigma_{0}^{2}\zeta \cos^{3}\theta -
(c^{2}+M)[\frac{\zeta^{2}\sigma_{0}}{A^{2} \cos\theta}\frac{\pi^{2}}{12}+\nonumber\\
&&l^{2}\zeta\sigma_{0}\cos\theta]+2 A b u \s_{0}\zeta\cos\theta+b^2\s_{0}\zeta\cos\theta)\label{energy},
\eea
shows the effect of loss in the exponential prefactor $e^G$.
\begin{figure}[h]
\begin{center}
\subfigure[$ $]{\label{lieb_bogo}}
  {\includegraphics[scale=0.6]{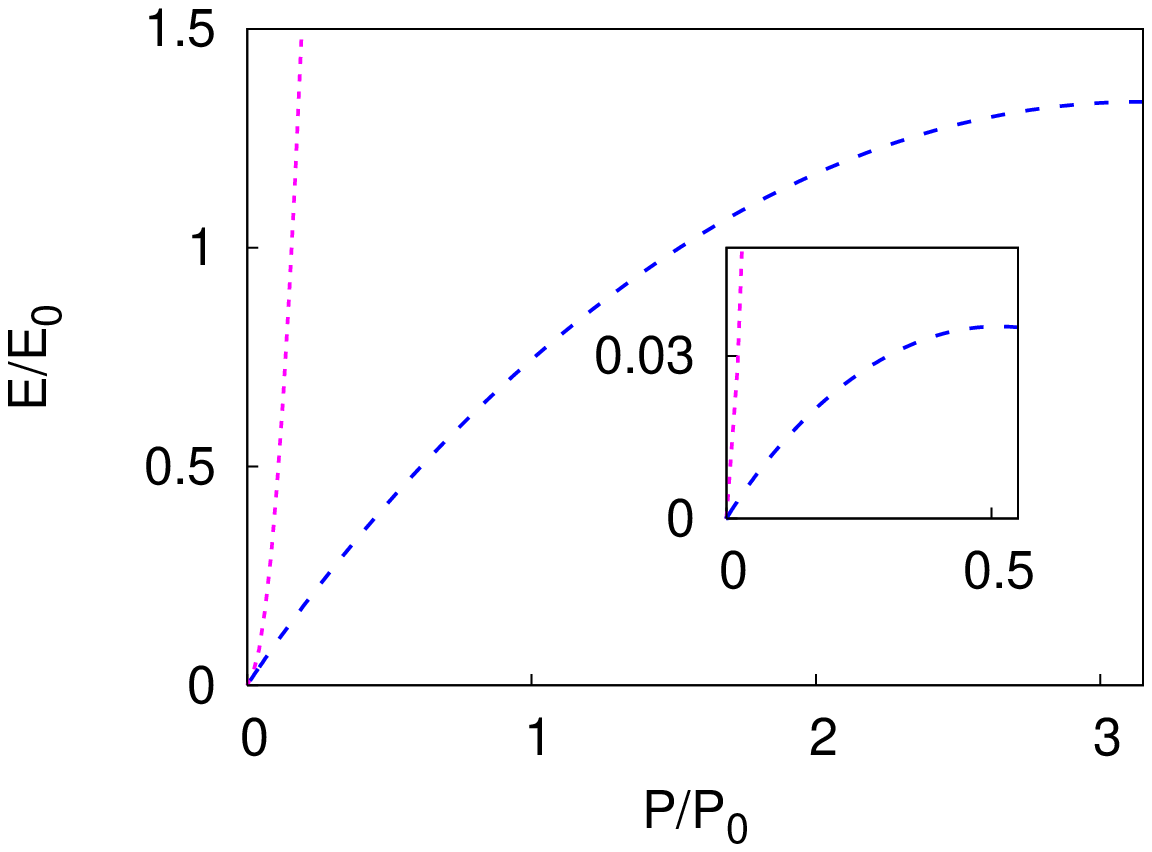}}
\subfigure[$ $]{\label{lieb}}
  {\includegraphics[scale=0.6]{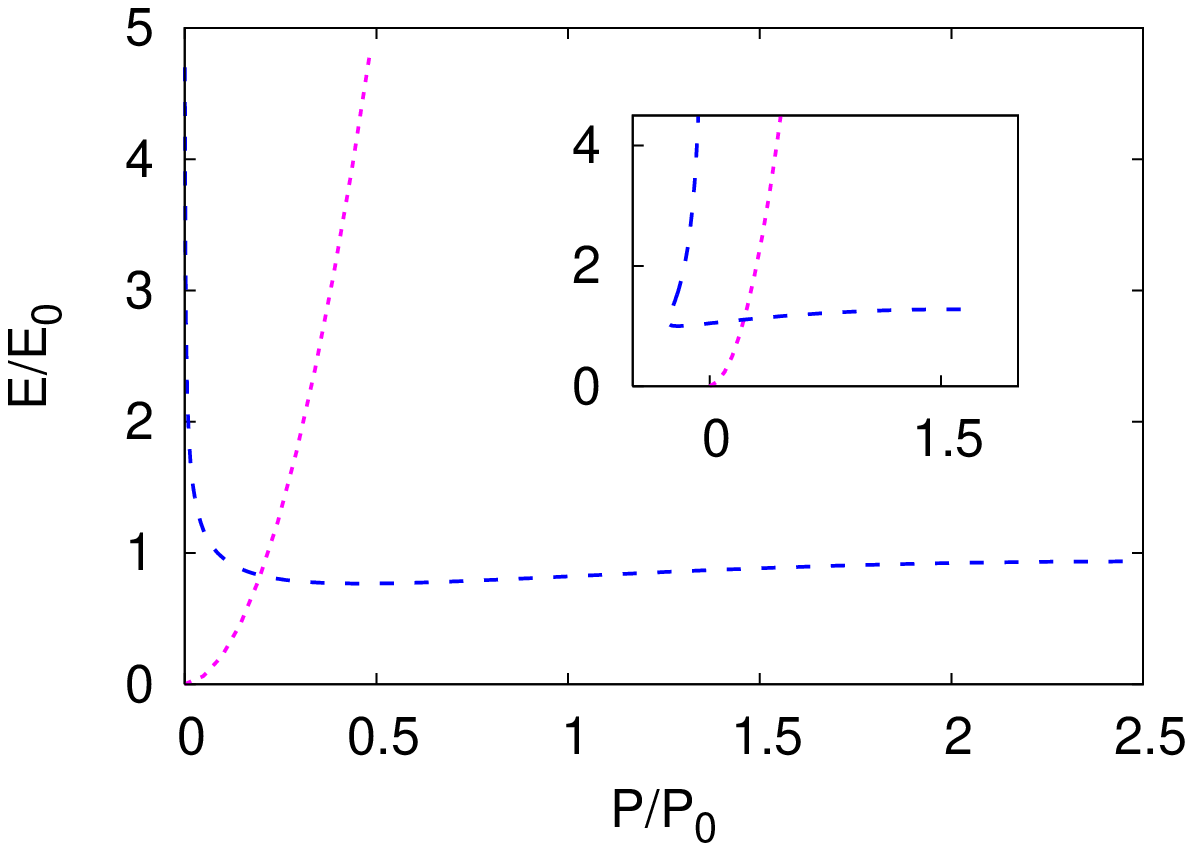}}
\caption{Dispersion relation for soliton (blue dashed line) and sound mode (pink dashed line). (a) The behavior of Lieb and Bogoliubov mode is shown with $2\pi$ periodicity at $t=0$. The inset depicts the dispersion relation without the trap at $t\neq0$. (b) The Lieb mode dispersion for $b=0$ inpresence of harmonic oscillator trap. In the inset depicts the mode behaviour when $b\neq0$.}
\end{center}
\end{figure}
The first term represents purely solitonic energy. The term thereafter accounts for the 
oscillator and chirping contributions, which is singular at $\theta=\pi/2$. This singularity arises from the physical consideration that, the trap does not favour uniform density distribution which excludes the asymptotic grey solitons, where $\s(z,t)=\s_{0}$. 
The third term, linear in $b(t)$, arises from the coupling of BEC momentum with soliton 
phase and the last term being quadratic in $b(t)$, represents BEC translational energy. 
Fig.\ref{lieb_bogo} depicts the well known $2\pi$ periodicity of the Lieb mode in absence of trap. In the inset it has been shown that if $t\neq 0$, the periodicity will be different. The energy and momentum is normalized by $E_{0}$ and $P_{0}$, where $E_{0}=\k\z(A_{0}\s_{0})^2$ and $P_{0}=A_{0}c_{s}\s_{0}\z$.
The strong effect of the trap at low momenta on the grey soliton's dispersion is shown in Fig.\ref{lieb}.
The presence of the translational motion of BEC affects soliton significantly and one observes that in certain cases at low momenta, the soliton's momentum can become negative, as depicted in the inset of Fig.\ref{lieb}. There is a point at which the energy is non zero at zero momentum, which reveals the rest mass of the solitons.

The canonical momentum is given by,
\bea
P&=&-i\int{\psi^{*}\frac{\partial \psi}{\partial z}}dz \nonumber\\
&=&e^{G} [C_{s}\zeta\sigma_{0} (\pi\frac{u}{|u|}-\sin 2\theta-2\theta)-
2b\zeta\sigma_{0}\cos\theta].
\eea
The first term can be attributed to soliton, and the second term arises from BEC momentum.
It is worth mentioning that the soliton velocity can also be computed from the hydrodynamic relation: $\frac{\partial E}{\partial P}=Au$, which matches with
the earlier obtained result.
In the limiting case, i.e., when the trap is switched off, all the expressions match with the known results \cite{jackson,komineas}.

In order to compare the behavior of the Lieb mode and the second sound, we apply a small perturbation \cite{barashenkov}:
\bea 
\delta\psi(z,t)&=&\sqrt{A(t)}\delta\rho(z,t)\exp[i\Phi(z,t)+G(t)/2],\nonumber\eea 
to a constant background $\psi_{0}$.
The modified equation of motion for $\delta\rho$ is given by, 
$i\partial_{t}\delta\rho=-\frac{1}{2}\partial_{zz}\delta\rho+\kappa A^2 [2|\rho_{0}|^2\delta\rho+\rho_{0}^2\delta\rho^{*}]-i\Phi_{z}\delta\rho$.
This can be written as an eigenvalue equation, $Jy_{t}=Hy$, 
where \begin{eqnarray*} J={\left (\begin{array}{cc}
0 & -1\\
1 & 0
\end{array}\right)}&,&
y={\left (\begin{array}{c}
\delta\rho_{R}\\
\delta\rho_{I}
\end{array}\right),}\end{eqnarray*}
and
\begin{eqnarray*}
H={\left (\begin{array}{cc}
-\frac{1}{2}\partial_{zz}+\kappa A^{2}\sigma_{0}
& -\Phi_{z}\partial_{z}\\ \Phi_{z}\partial_{z} & -\frac{1}{2}\partial_{zz}
\end{array}\right ).}\end{eqnarray*}
Assuming a plane wave perturbation i.e., $y(z,t)=\exp[i(kz-\omega t)]$, the dispersion relation can be obtained:
\be \omega=\sqrt{\frac{1}{4}k^{4}+C_{s}^{2}k^{2}}+(b-cz)k.\ee

The Bogoliubov dispersion carries the effect of BEC momentum and chirping in the second and third terms respectively.
The translational motion of the BEC modifies the Bogoliubov dispersion, as has been observed earlier in the 
case of BEC flowing with subsonic \cite{recati} and supersonic velocity \cite{recati,smerzi}. 
In the weak coupling regime, due to the presence of the trap, 
chirping is unavoidable. In the strong coupling scenario, Thomas-Fermi approximation accounts for the trap.
\section{Conclusion}
In conclusion, it is observed that, there is a nice and elegant way to extract analytical grey soliton solution from a rather complicated and more physically suited representation of the \GP equation.
Explicit calculations show that the harmonic confinement  significantly affects the energy momentum profile of the grey solitons at low momenta, since uniform density is not permitted in the trap. 
The Bogoliubov mode is weakly coupled with the trap. This opens up the possibility  for the coherent control and manipulation of these solitons at low energy, through the temporal modulation of the scattering length, as well as the trap. Explicit analysis of the center of mass motion shows that, the soliton can be compressed and accelerated through the trap and scattering length. We hope that, the strong coupling of the Lieb mode with the trap can be realised through the present experimental setups.

\section*{References}


\end{document}